# FACETS – a Framework for Parallel Coupling of Fusion Components

John R. Cary, Ammar Hakim, Mahmood Miah, Scott Kruger, Alexander Pletzer, Svetlana Shasharina, SrinathVadlamani
Tech-X Corporation,
Boulder USA

Alexei Pankin
Lehigh University,
Bethlehem, USA

Ronald Cohen, Tom Epperly, Tom Rognlien
Lawrence Livermore National Laboratory,
Livermore USA

Richard Groebner
General Atomics,
San Diego, USA

Satish Balay, Lois McInnes, Hong Zhang
Argonne National Laboratory,
Argonne, USA

*Abstract*— [1] Coupling separately developed codes offers an attractive method for increasing the accuracy and fidelity of the computational models. Examples include the earth sciences and fusion integrated modeling. This paper describes the Framework Application for Core-Edge Transport Simulations (FACETS).

I. INTRODUCTION

Computational efforts in the fusion and other communities have traditionally concentrated on solving physics models within distinct spatial regions, using approximations valid within well-defined ranges of temporal and spatial scales. This resulted in the development of numerous independent computational applications, each specializing in these different scales. Examples include radio-frequency waves (RF) propagation, for which the fundamental period is sub-nanosecond, magnetohydrodynamics (MHD), for which the period can range from microseconds to milliseconds, and gyrokinetic microturbulence (GK), for which time scales are sub-millisecond. At the other extreme of the time scale spectrum are transport calculations, which cover 1000 seconds or more for ITER. (The spatial scales are also substantially different, again much smaller for RF, MHD, and GK than for overall transport.)

This project was supported by DOE grants DE-FC02-07ER54907, DE-FG02-05ER84192 and Tech-X Corporation. This work performed under the auspices of the U.S. Department of Energy by Lawrence Livermore National Laboratory under Contract DE-AC52-07NA27344.

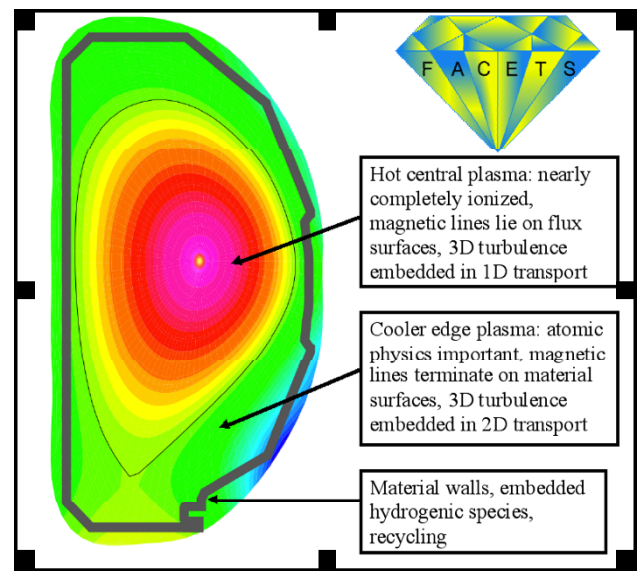

Figure 1. FACETS is about integrating core, edge and wall of tokamaks.

The problem of coupled core-edge transport simulations exemplifies the multiphysics challenges faced by the fusion program. The core and edge regions are very different in their spatial and temporal scales. Core plasma transport is dominated by turbulence with relatively short spatial scales. This transport can be summarized in terms of surface fluxes for the basic moments (densities, temperatures, and momenta) and so is essentially one-dimensional (radial). On the open field lines, which contact material walls, perpendicular and parallel transport compete, so that edge transport is two-dimensional and essentially kinetic. Thus, whole-device modeling requires the development of a multiphysics application able to use different computational approaches in different regions of the plasma.

The FACETS project [1-2] is the first significant effort aimed at combining the aspects of multi-physics and multiple computational regions (core, edge, and wall) within a single executable. It is designed to leverage the massively parallel computing resources available at supercomputing leadership class facilities and/or cover the multiple regions composing a tokamak, and by doing so enabling high fidelity integrated modeling simulations. Because integrated simulations aim to model the tokamak on a time scale much longer than many of the internal equilibration time scales, the framework must allow for implicit coupling.

The US Department of Energy, realizing the challenge of full-device and multiphysics modeling, has funded two other SciDAC integration projects, [3-5] that are addressing other computer science and physics aspects of coupled systems.

## II. FACETS REQUIREMENTS

FACETS software consists of a framework and utilities. The framework is the software that composes the fusion computational modules and advances them forward in time. Utilities support common build and test system, tools to standardize the components outputs, and tools to perform visualization.

The design of the framework and utilities has been driven by the requirements - some of which are common to all integrating efforts and some are specific for FACETS.

Common requirements include abilities to:

1) Incorporate legacy codes
2) Develop new fusion components
3) Use conceptually similar codes interchangeably
4) Incorporate components written in different languages

In addition FACET has adopted the requirements that the framework be able to:

5) Work well with the most rapid (simplest) computational models as well as be able to us the most computationally intensive models
6) Be applicable to implicit coupled-system advance
7) Take maximal advantage of parallelism by allowing concurrent execution

Requirement (5) leads to the need to provide tight coupling in order to accommodate the simplest computational components, which return values rapidly. Tight coupling means that components interact communicate in a synchronous and rapid (low latency) manner. This requires interaction through memory (rather then exchange files). In addition, tight coupling facilitates exception and error handing and lands better to the dynamic load balancing of components.

At the same time, to incorporate the more computationally demanding components, it is also desirable to keep the components in memory. As an alternative approach, one could have chosen to launch components using parallel job commands. However, this has a disadvantage of causing additional run-time overhead associated with memory allocation, data initialization, loading data from disk, and perhaps, the need to re-compute Jacobians.

Requirement (6), along with flexibility, requires a framework that allows runtime construction of algorithms, in order to be able to explore multiple coupling and time advance strategies. Furthermore, one must be able to recover from failures.

In addition, to provide better performance and further flexibility (compared to components granularity), FACETS is designed to have:

8) Construction process that allows for direct memory access
9) Separation of algorithms from data for algorithms reuse and aggregation
10) Flexible means for defining multiple types of integrated simulations without code recompilation
11) Run time discoverable implementations and instances

At present, only small amounts of data are being transferred between components – largely a few scalars and some small vectors. Hence there is presently little need for packing bulk data for data transfers in FACETS, as reflected by the current interface described below. Despite the low amount of data moved, the implicit coupling has required low latencies, and so the data transfer is performed directly from processor to processor without any intermediate processing using a single MPI Send/Recv.

In what follows we describe FACETS components, FACETS initialization process, composition language and support for parallel concurrent executions. We also describe the first coupling results and our next challenges.

## III. FACETS FRAMEWORK

### A. FACETS Components

At present, the FACETS framework has a component for core transport (FACETS::Core), neutral beam sources (NUBEAM), embedded turbulence (GYRO), and edge transport (UEDGE). In the near term, we will be incorporating a component for wall modeling (WALLPSI) and radiofrequency sources (TORIC). Here we define what is meant by a component, and discuss how we brought these components into our framework.

Conceptually, a FACET component is a unit of simulation which contains data representing its state and has agreed-upon interfaces. These interfaces define initialization and finalization, allocation of parallel resources, time update, data access and data output methods.

The whole initialization process (see Table 1) involves multiple steps. First, components set up logging files and set their MPI communicator. Then they allocate internal memory and build internal data structures. Next, they set up their algorithms by calling `buildUpdaters` function and initialize their fields by calling `initialize`. This function gets called at every new run and does not get called if the component restores itself from a stored state.

**Initialization/Finalization Interface**

```
int setLogFile(const string& lf);
int setMpiComm(long comm);
int readParams(const string& infile);
int buildData();
int buildUpdaters();
int initialize();
int finalize();
```

Table 1. Initialization/finalization interface of FACETS components.

The update interface (see Table 2) allows one to advance the state of a component in time as prescribed by their standalone physics behavior and to set and get data exchanged between components. If the advance of any component fails, such as due to solver non-convergence, a component can be reset to the last valid state.

**Update Interface**

```
int update(double t);
int revert();
```

Table 2. Update interface of FACETS components.

Data access interfaces (see Table 3) allow the setting and getting scalar and array data. These interfaces will become more general once FACETS starts using components that require higher dimensionality interfaces:

**Data Access Interface**

```
int getRankOfInterface(const string& name, size_t& ret)
int setDouble(const string& name, double val);
int getDouble(const string& name, double& ret) const;
int setDoubleAtIndex(const string& name, size_t ndims, const size_t[] indices, double val);
int getDoubleAtIndex(const string& name, size_t ndims, const size_t[] indices, double& ret) const;
int setDoubleAtLoc(const string& name, size_t ndims, const double[] loc, double val);
int getDoubleAtLoc(const string& name, size_t ndims, const double[] loc, double& ret) const;
```

Table 3. Data access interface of FACETS components.

Dump/restore interfaces allows dumping and restoring components from/to files and from/to particular file nodes:

**Dump/Restore Interface**

```
int dumpToNode(const string& file, const string& groupNode) const;
int dumpToFile(const std::string& file) const;
int restoreFromNode(const string& file, const string& groupNode);
int restoreFromFile(const std::string& file);
```

Table 4. Dump/restore interface of FACETS components.

All methods return an integer representing a code error (0 in case of no error) and pass a non-const reference to return an actual value for get methods. This choice of error handling is motivated by the need to support legacy codes that are written in Fortran, which does not support exceptions.

Components that represent codes developed independently from the FACETS framework are expected to implement the interfaces described above. In order to incorporate them into the framework, the FACETS team typically performs the following steps. First, one demands that the component developers provide or help generate a standalone test. Next the component is wrapped into the interfaces described above and a test of the wrapped component is included in the component's regression test system. The component is then brought into FACETS: meaning that a FACETS-style driver and FACETS-style input file is written for this component. This driver and input file are then added to the FACETS tests of regression tests and results are compared with the original standalone test. Finally, a test for the coupled system is developed to more rigorously test the interfaces and coupled system.

At present, we have performed this process for three standalone codes (UEDGE, NUBEAM, GYRO) and plan to do the same for three more codes in the coming year (WallPSI, BOUT++, and TORIC). The process is easily performed that codes whose interfaces can be described with the current interfaces. If the interfaces require higher dimensionality, then we would need more work, but we feel that the process that we currently use can be easily generalized.

Newly developed FACETS components (internal components) implement the above methods in a way that allows minimal indirect memory access, reuse of algorithms in a plug-and-play manner and use the extensive FACETS libraries for messaging, reading and writing data, grids, data structures, interpolation, component composition, etc.

FACETS enforces a separation of data from the algorithms that advance the data in time – *updaters* in the FACETS language. A FACETS updater is an abstraction of an algorithm or a function. Updaters have "in" and "out" data, which are specified by a name and a data type. The "in" data comes from components and the "out" data is put back in. The main difference between internal FACETS components and updaters is that updaters do not have an internal state, which is relevant to the simulation: they just perform the component data update. Such separation provides for more flexibility as updaters can be reused in multiple components.

Different updaters implement different algorithms and each component can have a complicated sequence of updaters. For example, the user can specify that the data structures representing the profiles of density and temperature be updated using a Crank-Nicholson algorithm or some other, non-time-centered algorithm. Another example is that the fluxes can be linearly combined from several independent flux calculations. Finally, the update step order may be used to combine the various updaters in a particular order. If any update returns a failure code, then all components are reset to their last valid step, and the step is retried with a smaller time step to automatically determine if the components' internal solvers are not converging due to an ill-conditioned advance.

Thus, the implementation of the `update` method of internal components is delegated to updaters, which contain direct references to the data structures of components and manipulate them to dynamically advance the state of a component. To provide this access, the `buildUpdaters` method (called after memory allocation of all objects, and before initialization or restoration of any object) implements direct access to the memory locations of the data of another object, so that at advance time, interface overhead disappears.

The coupling of components in FACETS is performed by container components that have internal updaters that describe the coupling algorithm. The updaters in the case take advantage of the component interfaces described previously. Each container component is responsible for assigning a subset of its processors to the contained components, while all its processors are available to all updaters as they need to communicate between data structures that can live on any processor of the container. Coupling order and frequency are defined in the input file and performed via special ranks of each component, which serve as points of the MPI send and receive commands.

The structure of the FACETS components hierarchy is shown on Figure 2. The base class `FcComponent` has the interface defined above. It is an abstract class. `FcContainer` derives from `FcComponent` and introduces the idea of hierarchical components: it can have a list of components within. `FcUpdaterComponent` introduces FACETS data structures and updaters. Components at this level delegate their update to the update methods of the updaters.

In order to be able to distinguish between different kinds of components, we also introduce abstract classes `FcWallIfc`, `FcCoreIfc`, and `FcEdgeIfc` (and more as needed). Each such class basically adds a class flavor that would allow imposing correct composition rules (for example, one can prohibit an edge component to contain a core component etc). These interfaces have methods that are very specific to the kinds they define. For example, `FcCoreIfc` should have the following methods setting and getting a particular variable (for example, `energyFlux_CE_electrons`) and add them into a map that would associate `setDouble` and `getDouble` methods with these functions:

```
FEdgeIfc : public FcComponent{
  FcEdgeIfc():FcComponent(){
    this->registerSetMethod
        ("energyFlux_CE_electrons",
         &FcEdgeIfc::setEnergyFlux);
    this->registerGetMethod
        ("energyFlux_CE_electrons ",
         &FcEdgeIfc::getEnergyFlux);
  }
  virtual double getEnergyFlux() = 0;
  virtual setEnergyFlux(const double) = 0;
};
```

Any external component would then have to derive from a particular interface (`FcWallIfc`, `FcCoreIfc` or `FcEdgeIfc`) and implement pure virtual functions specific for this kind of interface. For example, `FcExtEdgeComp` will derive from `FcEdgeIfc` and will have to implement `getEnergyFlux` and `setEnergyFlux` functions.

If the component is external (does not rely on FACETS updaters and data structures) this derivation is enough (for example, `FcExtEdgeComp` and `FcExtWallComp` on Figure 1). If a component uses FACETS infrastructure it will also have to derive from `FcUpdaterComponent` (for example, `FcCoreComp`). This inheritance introduces the diamond pattern, which is not dangerous as all inheritances are virtual. Examples of existing external components are `FcWallPsiComponent` and `FcUEDGEComponent`. An example of an internal FACETS component is `FcCoreComponent`.

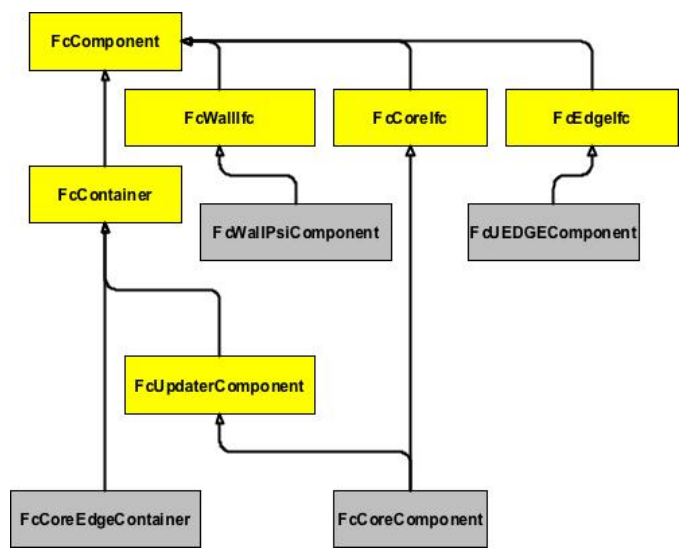

Figure 2. The schematic FACETS components hierarchy. Grey boxes show concrete components.

In order to tell container components what kind of containees they can have, we also introduce container classes. An example is the `FcCoreEdgeContainer` class, which expects exactly one contained edge and one contained core component if this type of container is specified.

*B. Input Language*

The FACETS input file describes the simulation composition. As always, the input file has to contain the parameters needed to describe the simulations. But for a flexible application as described above, the input file must also describe (1) the containment hierarchy, (2) the other objects needed a particular object for its update. For this purpose FACETS developed a simple XML-like language with allows certain tags, setting global constants and a means to describe numerical vectors.

The bulk of the input file is defining the simulation components (their data structures, grids and updaters). In addition, the input file describes how coupling is performed. For example, the following input file defines a coupled core-edge simulation. It starts with defining a top container component facets containing two members core and edge of kinds coreComponent and edgeComponent. Within each item, the input specifies components updaters and data structures being updated:

```
# This is a top container with core and edge
<Component facets>
  kind = coreEdgeContainer
# First child component
  <Component core>
    kind = coreComponent
# Data structs of core
    <DataStruct qOld>
      kind = distArray1D
      onGrid = coreGrid
    </DataStruct>
# Similar qNew is skipped
# Updater calculation qNew from qOld
    <Updater accept>
      kind = linCombinerUpdater
      in   = [qOld]
      out  = [qNew]
    </Updater>
# Manual load balancing
    load = 0.5
  </Component>
# Second child component
  <Component edge>
    kind = udgeComponent
# Manual load balancing
    load = 0.5
  </Component>
```

The keyword "load" instructs FACETS to manually allocate processor (in the case, equally between the core and edge components).

Coupling of components is performed by the updater, myCoreEdgeUpdater, which is of kind, explicitCoreEdgeUpdater. It specifies the names of the parameters, which are passed from one component to another:

```
# Updater coupling core and edge
  <Updater myCoreEdgeUpdater>
    kind = explicitCoreEdgeUpdater
    coreName = core
    edgeName = edge
# variables to pass from core to edge
    core2EdgeVars = ["heatFlux"]
# variables to pass from edge to core
    edge2CoreVars = ["temperature"]
  </Updater>
</Component>
```

By default, the data exchange is performed after every time step (once components advance by dt using their internal updaters not shown in the example).

Using coreEdgeContainer as a kind in the facets component enforces the framework to check for the presence of exactly one core and exactly one edge component within.

## C. Component Creation and Registries

For our dynamic discovery of components, we separate the concepts into the 3 I's: *Interface*, *Implementation*, and *Instantiation*. At startup, constructors for all implementations are stored in an associative array that allows one to construct such an object from a string. Thus, upon parsing the above file, and seeing that a component of kind extEdgeComponent is needed, FACETS does a lookup of the name, extEdgeComponent, and is returned a new instance of the class, FcExtEdgeComponent, which has the name, edge. That object is then given its section of the input file, which describes how that object will be constructed, and it is put into the component registry.

Having both implementation and instance registries provides flexibility. An implementation registry provides a mechanism discovery of available implementations. For example, a core source provides the power input to the core from some source, such as neutral beams or electromagnetic radiation (RF). As such, there are multiple implementations, including different implementations for the same physics but having varying degrees of fidelity and computational intensity.

## D. Support for Parallelism

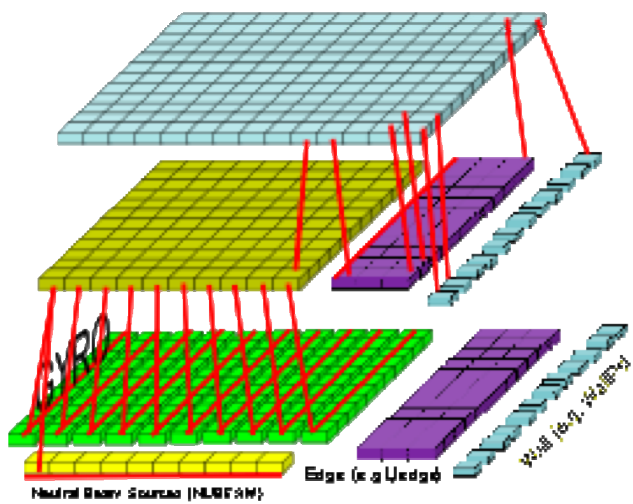

Figure 3. Processor breakup for a core-edge simulation with embedded turbulence.

FACETS provides several mechanisms to support parallelism. The simplest supported parallelism is domain decomposition, which is provided by a distributed array facility. Here, a N-dimensional array is split automatically over a given number of processors. This splitting is done using an algorithm that tries to assign equal volume of data to each processor, making the computational load on each processor approximately equal. However, no attempt is made to equalize the amount of data communicated across processors.

A more sophisticated form of domain decomposition is also supported for use in embedded turbulence calculations. For such problems a set of transport equations is solved on a one-dimensional domain, the transport fluxes obtained at each time-step from a local turbulence calculation. Each turbulence calculation on a flux surface itself is very computationally intensive and runs in parallel. Once it is complete, the computed fluxes are communicated to the transport solver and the solution is advanced. Hence, a special form of decomposition, in which each flux surface is assigned many processors, is needed.

This decomposition is done by introducing the concept of multi-processor arrays. These arrays are created as follows. First, a large set of processors is reserved for the transport solver, which runs on, for example, a one-dimensional domain divided into 32 cells. These processors are divided into 32 sets of processors. Each of these sets is used in the turbulence flux calculations itself. In addition, to allow for communication of the fluxes back to the transport solver, the zero-rank of each set is assembled into a communicator on which a distributed array is allocated. This allows the communication of gradients and values to the turbulence flux calculators and the fluxes back to the transport solvers.

Figure 3 shows a planned processor distribution for a FACETS simulation with a core component consisting of embedded turbulence and beam sources, an edge component and a wall component consisting of multiple wall segments. In the first step, all available processors are split between the core, edge and wall components. Then, in the core component the processors are further split between the beam sources, itself a Monte-Carlo task parallel component, and the turbulence calculators using the multi-proc array facility described above. The edge component splits up its share of processors in a traditional domain decomposition fashion. The wall component splits its share of processors among the wall segments, each processor handling a set of wall tiles. Inter-component communication is handled by the framework by exchanging surfacial data needed for the coupling algorithm.

## IV. FACETS UTILITIES

### A. Language Interoperability

FACETS is using F90 modules representing turbulent transport models such as glf23 and mmm95. In addition, it brings in F90/Python codes such as UEDGE [6] and C-based WallPSI. These codes are rewritten as libraries with several methods exposed and wrapped into Babel's SIDL [7] so that they can be called from the C++ FACETS code.

### B. Build System and Regression Tests

FACETS development environment imposes strict discipline for individual developers. Prior to committing new code to the SVN repository, one has to run the full set of FACETS tests (fctests). They report violation in coding standards (formatting, documenting, layering rules) as well as failure to build or differences in numerical results. In addition, these tests are run nightly and notify the team about the results by an email.

### C. Standard output and visualization

Validation and Verification efforts require a standard output format. All FACETS components abide to the VizSchema [8-9] standard, which uses HDF5 as its underlying file format. Therefore, FACETS output data are portable across platforms, including from 32 to 64 bit architectures. The HDF5 API supports parallel write operations.

In addition to raw data, VizSchema also stores descriptive metadata such as field names, time slice information, grid resolution, etc. This enables any postprocessing application to manipulate FACETS output data without the need for additional, built-in knowledge. Additional markups/metadata were developed to identify variables living on different domains but which are conceptually the same. An example would be the temperature field, which extends from the core region to the edge. Establishing this connection is critical in order to visualize multi-grid seamlessly data across the entire domain.

Based on the VizSchema standard, we developed a data-reading module for the VisIt visualization tool. Figure 4 shows visualization of electron temperature coming from four different regions of FACETS simulation: three coming from an edge component and one from a core component.

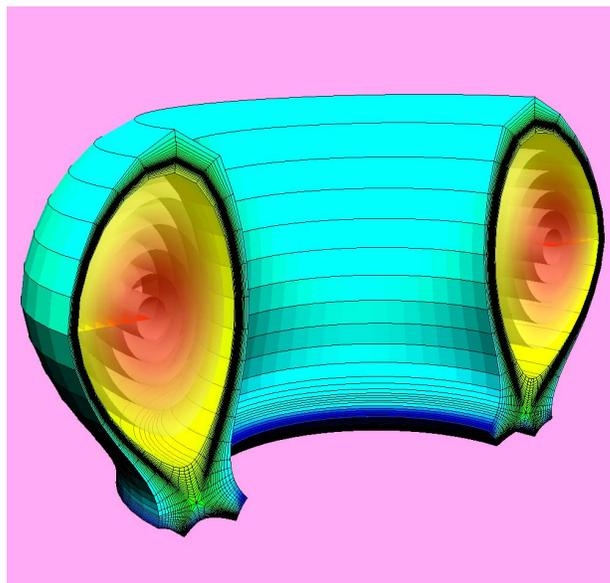

Figure 4. Multi-domain visualization of electron temperature coming from four regions from two FACETS components.

## V. FIRST COUPLING RESULTS

We have performed the first coupled core-edge simulations to validate FACETS solvers against experimental data. In these simulations we have initialized the core and the edge components from experimental data obtained from shot 118897 on the DIII-D tokamak at General Atomics Corporation. The core component advances the one-dimensional transport equations using a combination of fluxes from a reduced turbulence model, a neo-classical transport model and a combination of sources from beam heating, Ohmic heating and inter-species equilibration. The edge component advances the two-dimensional fluid equations using diffusivities that are constant in time but spatially varying to create a transport barrier near the separatrix. The components are coupled using an explicit coupling scheme. In this scheme, the core and the edge components are each advanced by a time-step and the fluxes at the core-edge interface sent from the core to the edge and the temperatures sent from the edge to the core. Using these new boundary values the next time-step is executed.

Figure 5 shows results of the evolution of ion temperature profiles from a FACETS simulation. The heavy cyan line represents the initial experimental data and the heavy black line represents the final experimental data, at the end of the shot. The various other lines are FACETS obtained profiles. It is observed that the ion temperature rises as seen in the experiment and the agreement of the final profile is reasonable given the simple diffusivity model in the edge and the fact that the densities were not evolved in the simulation.

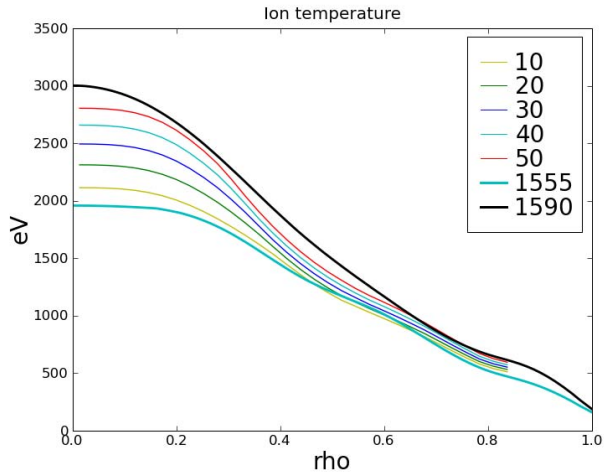

Figure 5. Ion profile evolution in a core-edge integrated simulation. Heavy black line is the experimental profile at the end of the shot. The red line is the profile at the end of the FACETS simulation.

VI. FUTURE DIRECTIONS FOR FRAMEWORK DEVELOPMENT

At present, we have incorporated the standalone codes UEDGE, NUBEAM, and GYRO. Our current physics studies are focused on improving the core-edge simulations described previously to understand neutral fueling of the experimental discharge along with the most important features of core-edge modeling. We are also investigating high-fidelity core transport simulations using embedded turbulence (GYRO) and Fokker-Planck modeling of neutral beam sources (NUBEAM).

Manually assigning computer resources to each component will become impractical as the complexity of FACETS, and the number of components, continues to grow. In order to achieve high scalability on Leadership Class Facility (LCF) computers with 10,000-100,000 cores or more, there is a need to automatically, and optimally spread processing power across components.

Our first approach will be to initially (statically) assign processing resources to each component through a negotiation process implemented in FACETS. The number of processors assigned to each component will depend on the total availability of processors, the demands of other components, and their intrinsic scalability characteristic. Processor allocation will also need to take into account the restrictions imposed by the components. For instance, some components may only be capable or running on specific number of processors. This approach will require extensions to our input language vocabulary, which presently uses a single "load" parameter to specify the allocation of processors relative to the total number of available processors.

We also recognize that the above approach, while being a significant improvement over manual load balancing, has some shortcomings. In particular, to work effectively, it relies on the scalability information of each component, which can be inaccurate and/or be highly dependent on specific hardware or simulation conditions. Thus, the execution time of an implicit solver advancing plasma profiles may be highly sensitive to the prevailing plasma conditions, which change from time slice to time slice. Therefore, we also anticipate the need at some point for dynamic load balancing. Here, we have much to learn from computer science where similar challenges are encountered. Consider for instance web servers and multi-tasking in modern operating systems, which distribute system resources to tasks, typically based on a first-come-first-served scheduling algorithm. However, it is important to recognize that in contrast to computer science use-cases, FACETS tasks are tightly coupled, with information being exchanged across tasks and synchronization barriers preventing the independent execution of tasks in arbitrary order. Therefore, we regard this as being one of our greatest challenges.

In addition to attention to dynamic balancing, the FACETS teams started collaboration with the SWIM and CPES teams to identify the particular niches that each of the projects addresses and come up with a unified plan to face the challenge of the integrated modeling.

VII. SUMMARY AND COMPARISON WITH OTHER APPROACHES

FACETS provides a flexible infrastructure for creating tightly coupled parallel simulations. Its approach is to incorporate legacy components by requiring a standard wrapper for communication and a means to send and receive MPI commands. Simulation is composed from an input file and does not have to be recompiled for different configurations. Newly developed FACETS components separate their state and updating mechanism and delegate their update methods to updaters. This separation promotes flexibility and code reuse.

FACETS is ported to multiple high-performance computers including LCFs and allows components written in different programming languages. FACETS stresses the need for standardization including standard output format and standard build system for all the components.

Other comparable fusion projects like SWIM and CPES are somewhat distinct from FACETS, although the differences and commonalities should be understood to move together to the full device modeling in the future.

SWIM's [3] framework called Integrated Plasma Simulator (IPS) is a Python based system which provides very light Python wrappers around legacy components for a lean interface allowing starting, running and finalizing the components. IPS has a set of common services (component registration and monitoring, for example). Each component runs its simulation to completion, dumps data in files in various formats, one of which represents to complete plasma state, which then can be used by other components. Thus, the system is well suited for loose coupling. The benefit of this approach is leaving physics codes untouched and using approaches familiar to physicists so it will be attractive to many modelers. The possible drawback of the approach is using file communication through a "bag" of data, which can be overwritten inconsistently and might not scale for a large number of parallel components.

CPES's [5] framework called End-to-End Framework for Fusion Integrated Simulations (EFFIS) pays a special attention to workflows and treats coupled simulations as such. To orchestrate the integrated simulations, EFFIS use the Kepler [10] workflow engine. Another distinct feature of EFFIS is unified approach to components I/O: each component uses ADIOS [11] for its output, which allows hiding the differences between the output data formats and choosing the underlying I/O mechanism suitable for the used platform. Until recently, data between components was exchanged using files, but the recent advances indicate that EFFIS is moving to in-memory coupling using Remote Direct Memory Access (RDMA). The possible drawback of EFFIS is dependence on the Java-based Kepler, which might present a problem on some supercomputers. Another problem could the use of RDMA, which is less familiar to computational scientists than MPI.

ACKNOWLEDGMENT

The authors thank the whole FACETS team and DOE SCiDAC program.